\documentclass[10pt,twocolumn]{IEEEtran}
\usepackage{amsfonts,dsfont}
\usepackage[tbtags]{amsmath}
\usepackage{graphicx,subfigure}
\usepackage{color}      
\usepackage{epsfig}
\usepackage{amssymb}
\usepackage{tipa}
\usepackage{bbm}

\long\def\comment#1{}

\newcommand{\beq}{\begin{equation}}
\newcommand{\eeq}{\end{equation}}
\newcommand{\beqno}{\begin{equation*}}
\newcommand{\eeqno}{\end{equation*}}

\newcommand{\bes}{\begin{split}}
\newcommand{\ees}{\end{split}}
\newcommand{\bdm}{\begin{displaymath}}
\newcommand{\edm}{\end{displaymath}}

\newtheorem{definition}{Definition}

\newcommand{\bd}{\begin{definition}}
\newcommand{\ed}{\end{definition}}

\newcommand{\bv}{\begin{vugraph}}
\newcommand{\ev}{\end{vugraph}}
\newcommand{\bi}{\begin{itemize}}
\newcommand{\ei}{\end{itemize}}
\newcommand{\ben}{\begin{enumerate}}
\newcommand{\een}{\end{enumerate}}

\newcommand{\bean}{\begin{eqnarray*} }
\newcommand{\eean}{\end{eqnarray*} }
\newcommand{\bea}{\begin{eqnarray} }
\newcommand{\eea}{\end{eqnarray} }
\newcommand{\nn}{\nonumber}
\newcommand{\ba}{\begin{array} }
\newcommand{\ea}{\end{array} }

\newcommand{\n}{{\cal N}}

\begin{document}
\renewcommand{\textfraction}{0}

\title{Performance evaluation for ML sequence detection in ISI channels with Gauss Markov Noise}

\author{\authorblockN{Naveen Kumar\authorrefmark{1}, Aditya Ramamoorthy\authorrefmark{1} and Murti Salapaka\authorrefmark{2}} \\\authorblockA{\authorrefmark{1}Dept. of
Electrical and Computer Engineering\\ Iowa State University,
Ames, IA 50010\\
Email: nk3,adityar@iastate.edu}
\\ \authorblockA{\authorrefmark{2}Dept. of Electrical and Computer Engineering\\University of Minnesota, Minneapolis, MN 55455
\\Email: murtis@umn.edu} }

\maketitle

\begin{abstract}
Inter-symbol interference (ISI) channels with data dependent Gauss
Markov noise have been used to model read channels in magnetic
recording and other data storage systems. The Viterbi algorithm can
be adapted for performing maximum likelihood sequence detection in
such channels. However, the problem of finding an analytical upper
bound on the bit error rate of the Viterbi detector in this case has
not been fully investigated. Current techniques rely on an
exhaustive enumeration of short error events and determine the BER
using a union bound.

In this work, we consider a subset of the class of ISI channels with
data dependent Gauss-Markov noise. We derive an upper bound on the
pairwise error probability (PEP) between the transmitted bit
sequence and the decoded bit sequence that can be expressed as a
product of functions depending on current and previous states in the
(incorrect) decoded sequence and the (correct) transmitted sequence.
In general, the PEP is asymmetric. The average BER over all possible
bit sequences is then determined using a pairwise state diagram.
Simulations results which corroborate the analysis of upper bound,
demonstrate that analytic bound on BER is tight in high SNR regime.
In the high SNR regime, our proposed upper bound obviates the need
for computationally expensive simulation.

\end{abstract}

\vspace{-.3cm}
\section{Introduction}
Maximum likelihood sequence detection (MLSD) in channels with
inter-symbol-interference and data dependent time-correlated noise
is an important problem in many domains. For example, in magnetic
recording, the statistics of percolation and nonlinear effects
between transitions \cite{noisewang95,kavcic98} result in noise that
exhibits data-dependent time-correlation. Recently, similar noise
models for nanotechnology based probe storage have also been
developed and the corresponding detectors have been found to have
significantly improved performance compared to the current state of
the art \cite{kumar2010}. It is well-recognized that a sequence
detector designed for an AWGN ISI model can have a significant loss
of performance if the data dependence and time-correlation of the
noise is not taken into account.

In the case of finite ISI channels with memoryless noise, Forney
\cite{forney1972, forney73viterbi} presented an MLSD solution based
on the Viterbi algorithm. Upper bounds on the error probability of
the detector can be derived based on flowgraph techniques
\cite{biglieri84, biglieri90, omura79}. The work of Kavcic \& Moura
\cite{kavcic2000} considered channels with finite ISI and noise
modeled by a finite memory Gauss-Markov process. The work of
\cite{kavcic2000}, also presents certain approaches (see section
\ref{sec:conclusions} in \cite{kavcic2000}) for computing an upper
bound on the performance of the detector. However, their technique
is not based on flowgraph techniques, and requires an enumeration of
all error events of relevant lengths and an estimate of the
corresponding pairwise error probability upper bound. We emphasize
that an analytical technique for estimating detector performance is
of great value since it allows us to predict the performance at high
SNR's where simulation can be time-consuming.

In an ISI channel with additive white Gaussian noise (AWGN), the
pairwise error probability (PEP) between two state sequences can be
easily factorized as a product of functions depending on current and
previous states in the (incorrect) decoded sequence and the
(correct) transmitted sequence. Let $\bar{S}$ and $\hat{\bar{S}}$ be
the transmitted and decoded state sequences respectively. Then this
means that the probability that the detector prefers $\hat{\bar{S}}$
to $\bar{S}$, is denoted by
$P(\hat{\bar{S}}|\bar{S})=\Pi_{k=0}^{N-1}
h(\hat{S}_{k-1}^k,S_{k-1}^k)$ where $h$ is a function of current
state and previous decoded states $\hat{S}_{k-1}^k = (\hat{S}_{k-1},
\hat{S}_k)$ and actual states $S_{k-1}^k = (S_{k-1}, S_k)$.
Moreover, the PEP is symmetric due to the symmetric nature of white
Gaussian noise , i.e., $P(\hat{\bar{S}}|\bar{S}) =
P(\bar{S}|\hat{\bar{S}})$. Together, these properties allow the
application of the error state diagram method for finding an upper
bound on the BER \cite{omura79}.

In contrast, for the ISI channel with data-dependent Gauss-Markov
noise (considered in \cite{kavcic2000}), neither of these properties
hold. The signal dependent and time-correlated noise makes the PEP
asymmetric. Further the PEP does not factorize in a suitable manner
as required for the application of flowgraph techniques. This makes
the estimation of BER for such channels, quite challenging.

\noindent \underline{{\it Main Contributions:}} In this paper, we
consider a subset of the class of ISI channels with Gauss-Markov
noise. For these channels, we arrive at an upper bound to the PEP
that can be expressed as a product of functions depending on current
and previous states in the (incorrect) decoded sequence and the
(correct) transmitted sequence. The asymmetric character of the PEP,
i.e., the fact that $P(\hat{\bar{S}}|\bar{S}) \neq
P(\bar{S}|\hat{\bar{S}})$ necessitates an average over all correct
and erroneous state sequences. We show that this can be achieved
using the concept of the ``pairwise state diagram"
\cite{biglieri84}. Based on this, we present an analytical technique
for determining an upper bound on the BER. Simulations results show
that our proposed bound is tight in the high SNR regime.


The paper is organized as follows.
Section~\ref{sec:channel_modelling} introduces the channel model and
describes the corresponding Viterbi decoding algorithm.
Section~\ref{sec:upper bound} presents an upper bound on the
detector BER. Section~\ref{sec:Simulation Results} demonstrates
simulation results that confirm the analytical bounds.
Section~\ref{sec:conclusions} summarizes the main findings of this
paper and outlines future work.

\section{Channel model and Viterbi detector}\label{sec:channel_modelling}
We introduce the channel model and the corresponding detector in
this section. A word about notation. In what follows, if $z_k$ is a
discrete-time indexed sequence at k$^{th}$ time instant, the column
vector of sequence samples from time instant $k_1$ to $k_2$ is
denoted by $z_{k1}^{k2} = [z_{k1} \dots z_{k2}]^T$ where $k_1 \le
k_2$. We will use the notation $f(\cdot|\cdot)$ to denote a
conditional pdf. The precise pdf under consideration will be evident
from the context of the discussion.
\subsection{Channel Model}
\begin{figure}
\centering
\includegraphics[width=3.5in]{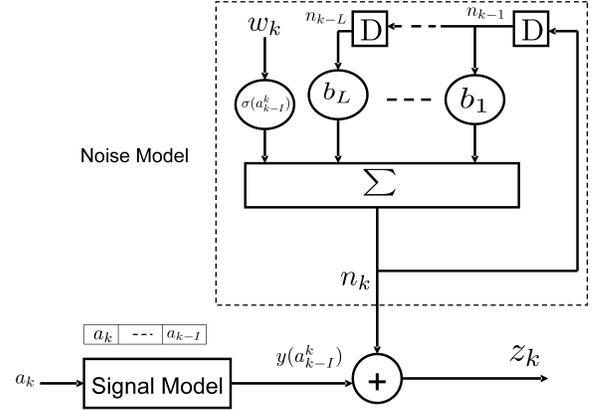}

\caption{\label{fig:channel_model}  Channel model with Gauss-Markov
noise.}

\end{figure}

Let $a_k$ denote the $k^{th}$ source bit that is equally likely to
be $0$ or $1$. The channel output shown in
Figure~\ref{fig:channel_model} with intersymbol interference (ISI)
of length $I$ is given by,
\begin{eqnarray*}
    z_k = y(a_{k-I}^{k}) + n_k,
\end{eqnarray*}
where $y(a_{k-I}^{k})$ is the noiseless channel output dependent
only on the $I+1$ past transmitted bits. The noise $n_k$ is modeled
as a signal dependent Gauss-Markov noise process with memory length
$L$ as explained below.
\begin{eqnarray*}
    n_k = \bar{b}  ^T n_{k-L}^{k-1}+ \sigma(a_{k-I}^{k}) w_k,
\end{eqnarray*}
where the vector $\bar{b}$ represents $L$ coefficients of an
autoregressive filter, $\sigma(a_{k-I}^{k})$ is signal dependent
parameters and $w_k$ is a zero mean unit variance i.i.d Gaussian
random variable. Note that in the most general model (considered in
\cite{kavcic2000}), even the autoregressive filter $\bar{b}$ would
depend on the data sequence $\bar{a}$. However, in this work, we
only work with models where $\bar{b}$ is fixed. We revisit this
point in Section \ref{sec:upper bound}. The noise $n_k$ can be
rewritten as,
    \begin{eqnarray*}
    n_k &=& \bar{b}  ^T \begin{pmatrix}n_{k-L}\\.\\.\\.\\n_{k-1} \end{pmatrix} + \sigma(a_{k-I}^{k})
    w_k\\
    &=& \bar{b}  ^T \begin{pmatrix}z_{k-L}- y(a_{k-L-I}^{k-L})\\.\\.\\.\\z_{k-1}- y(a_{k-1-I}^{k-1}) \end{pmatrix} + \sigma(a_{k-I}^{k})
    w_k
\end{eqnarray*}

This implies that
\begin{equation*}
    z_k = y(a_{k-I}^{k}) +\bar{b} ^T \begin{pmatrix}z_{k-L}- y(a_{k-L-I}^{k-L})\\.\\.\\.\\z_{k-1}- y(a_{k-1-I}^{k-1}) \end{pmatrix} + \sigma(a_{k-I}^{k}) w_k
\end{equation*}
From above analysis, we can conclude that
\begin{eqnarray}\label{eq:pdf1}
 f(z_k|z_0^{k-1},\bar{a}
)=f(z_k|z_{k-L}^{k-1}, a_{k-L-I}^{k} ),
\end{eqnarray}
where we recall that $f(\cdot|\cdot)$ represents the conditional
pdf.

\subsection[Viterbi Detector]{Viterbi Detector}

The maximum likelihood estimate of the bit sequence denoted
$\hat{\bar{a}}$ is given by
\begin{align*} & \hat{\bar{a}} = \arg
\max_{\bar{a}\in \{0,1\}^N} ~ f( \bar{z}|\bar{a} )\\
&=\arg \max_{\bar{a}\in \{0,1\}^N} ~ \Pi_{k=0}^{N-1}
f(z_k|z_0^{k-1},\bar{a} )\\
&=\arg \max_{\bar{a}\in \{0,1\}^N} ~ \Pi_{k=0}^{N-1}
f(z_k|z_{k-L}^{k-1}, a_{k-L-I}^{k} )\mbox{~(Using (\ref{eq:pdf1}))}\\
&=\arg \max_{\bar{a}\in \{0,1\}^N} ~ \Pi_{k=0}^{N-1}
\frac{f(z_{k-L}^{k}| a_{k-L-I}^{k} )}{f(z_{k-L}^{k-1}| a_{k-L-I}^{k}
)}
\end{align*}
We define a state $S_k= a_{k-L-I+1}^k$ (there will
   be a total of $2^{L+I}$ states). With this definition, $f(z_{k-L}^{k}| a_{k-L-I}^{k})=f(z_{k-L}^{k}|
   S_{k-1}^{k})$. Moreover it is Gaussian distributed,
   \begin{eqnarray*}
f(z_{k-L}^{k}|
   S_{k-1}^{k}) \sim N(\bar{\mu}(S_{k-1}^k), C(S_{k-1}^k))
   \end{eqnarray*}
where $\bar{\mu}(S_{k-1}^k)$ is the mean vector and $C(S_{k-1}^k)$
is the covariance matrix.

With our state definition, we can reformulate the detection problem
as the following MLSD problem.
\begin{eqnarray*} \hat{\bar{S}} &=&\arg \max_{\bar{S}}
~ \Pi_{k=0}^{N-1} \frac{f(z_{k-L}^{k}|
a_{k-L-I}^{k} )}{f(z_{k-L}^{k-1}| a_{k-L-I}^{k} )}\\
&=& \arg \max_{\bar{S}} ~ \Pi_{k=0}^{N-1}
\frac{f(z_{k-L}^{k}| S_{k-1}^k )}{f(z_{k-L}^{k-1}| S_{k-1}^k )}\\
&=& \arg \min_{\bar{S}} ~ \sum_{k=0}^{N-1} \log
\frac{|C(S_{k-1}^k)|}{|{\bf{c}}(S_{k-1}^k)|}\\&+&
(z_{k-L}^k-\bar{\mu}(S_{k-1}^k))^T
C(S_{k-1}^k)^{-1}(z_{k-L}^k-\bar{\mu}(S_{k-1}^k))\\&-&
(z_{k-L}^{k-1}-\bar{\mu}'(S_{k-1}^k))^T
{\bf{c}}(S_{k-1}^k)^{-1}(z_{k-L}^{k-1}-\bar{\mu}'(S_{k-1}^k))
\end{eqnarray*}
where $\hat{\bar{S}} $ is the estimated state sequence,
${\bf{c}}(S_{k-1}^k)$ is the upper $L \times L $ principal minor of
$C(S_{k-1}^k)$ and $\bar{\mu}'(S_{k-1}^k)$ collects the first $L$
elements of $\bar{\mu}(S_{k-1}^k)$. It is assumed that the first
state is known. With the metric given above, Viterbi decoding can be
applied to get the ML state sequence \cite{viterbi67} and the
corresponding bit sequence.

The matrix $C(S_{k-1}^k)$ is of dimension $(L+1) \times (L+1)$. For
higher values of $L$, the complexity of detector increases as the
decoding metric involves the inversion of the matrix $C(S_{k-1}^k)$.
However, the matrix inversion lemma can be used here to obtain
\begin{eqnarray}
 &&C(S_{k-1}^k)^{-1}=\left[ \begin{array}{cc}
{\bf{c}}(S_{k-1}^k) & \bar{c}\nn \\ \label{eq:matrixlemma} \bar{c}^T
& c
\end{array} \right]^{-1} \\&&= \left[ \begin{array}{cc}
{\bf{c}}(S_{k-1}^k)^{-1} & 0\\
0 & 0
\end{array} \right] + \frac{\bar{w}(S_{k-1}^k)\bar{w}(S_{k-1}^k)^T}{\gamma(S_{k-1}^k)},\end{eqnarray}
\noindent where
\begin{equation*}
\bar{w}(S_{k-1}^k)=\left[ \begin{array}{c}
-{\bf{c}}(S_{k-1}^k)^{-1}\bar{c}\\
1
\end{array} \right]=\left[ \begin{array}{c}
-\bar{b}\\
1
\end{array} \right], \text{~and}
\end{equation*}

\begin{equation*}
\gamma(S_{k-1}^k) =  (c- \bar{c}^T
{{\bf{c}}(S_{k-1}^k)}^{-1}\bar{c})= \sigma^2(a_{k-I}^k).
\end{equation*}

Using (\ref{eq:matrixlemma}), we can simplify the detector as
follows.


\begin{eqnarray*} \hat{\bar{S}}
&=& \arg \min_{\bar{S}} ~ \sum_{k=0}^{N-1} \log \sigma^2(a_{k-I}^k)
\\&&+ \frac{([-\bar{b}^T ~1
](z_{k-L}^k-\bar{\mu}(S_{k-1}^k)))^2}{\sigma^2(a_{k-I}^k)}.
\end{eqnarray*}
It should be noted that the above expression does not involve any
matrix inversion. This reduces the complexity of the detector
substantially. Another observation is that the Viterbi decoding
metric involves passing $z_{k-L}^k$ through a filter $[-\bar{b}^T ~1
]$ which is the inverse of the autoregressive filter of noise
process $n_k$ shown in Figure \ref{fig:channel_model}. The metric
first uncorrelates the noise with an FIR filter and then applies the
Euclidean metric to the output of
the filter. 

\section{Upper Bound on BER}\label{sec:upper bound}

As discussed previously, the channel model under consideration (cf.
Section \ref{sec:channel_modelling}), is such that the corresponding
PEP is asymmetric, and moreover does not factorize as a product of
appropriate functions as required by flowgraph techniques. We now
show that we can address this issue by using the Gallager upper
bounding technique \cite{gallager68}, coupled with a suitable change
of variables.

Denote an error event of length $N$ as
$\epsilon_N=(\bar{S},\hat{\bar{S}})$ such that $\bar{S}$ and
$\hat{\bar{S}}$ are valid state sequences and $S_k=\hat{S}_k$,
$S_{k+N}=\hat{S}_{k+N}$, $S_{k+j} \ne \hat{S}_{k+j}$ for $1 \le j
\le N-1$ and  $S_{k+j} = \hat{S}_{k+j}$ for other values of $j$
where $\hat{S}_k$ and $S_k$ are the estimated and correct state
respectively. Using this, an upper bound on the BER can be found as
follows \cite{biglieri84},
\begin{align*}
&P_b(e) \le \sum_{N=1}^{\infty} \sum_{\bar{S}} P (\bar{S})
\sum_{\hat{\bar{S}}: ~(\bar{S},\hat{\bar{S}}) \in E_N }
\nu(\bar{S},\hat{\bar{S}}) P(\hat{\bar{S}}|\bar{S}),
\end{align*}
where $\nu(\bar{S},\hat{\bar{S}})$ is the number of erroneous bits
along the sequences $\bar{S}$ and $\hat{\bar{S}}$ and $E_N$ is the
set of all error events $\epsilon_N$ of length $N$. The number of
erroneous bits is given by
\begin{align*}
&\nu(\bar{S},\hat{\bar{S}})=\frac{d}{dZ} \big{[} \Pi_{k=0}^{N-1}
Z^{\delta(a_k,\hat{a}_k)} \big{]}\big{|}_{Z=1}
\end{align*}
where $\delta(a_k,\hat{a}_k)=1,~\text{if}~ a_k \ne \hat{a}_k$ and
$Z$ is a dummy variable. Using this the upper bound above can be
expressed as
\begin{align*}
P_b(e) &\le \sum_{N=1}^{\infty} \sum_{\bar{S}} P (\bar{S})
\sum_{\hat{\bar{S}}: ~(\bar{S},\hat{\bar{S}}) \in E_N }
\frac{d}{dZ}\\ &\cdot \big{[}  \Pi_{k=0}^{N-1}
Z^{\delta(a_k,\hat{a}_k)}|_{Z=1} P(\hat{\bar{S}}|\bar{S})\big{]}
\end{align*}
where $P (\bar{S})= P(S_0)P(S_1|S_0)\dots
P(S_N|S_{N-1})=\frac{1}{M}.\frac{1}{2^{N}}$ if $\bar{S}$ is valid
state sequence , ($M$ is the number of states). The upper bound on
the PEP can be using Gallager's technique \cite{gallager68} as shown
below. Let $A(\bar{S}, \hat{\bar{S}}) = \{\bar{z} :
f(\bar{z}|\hat{\bar{S}}) \ge f(\bar{z}|\bar{S})\}$. Note that using
previous arguments, we also have that $A(\bar{S}, \hat{\bar{S}}) =
\bigg{\{}\bar{z} :
\Pi_{k=0}^{N-1}\frac{f(z_k|\hat{S}_{k-1}^{k},z_{k-L}^{k-1})}{f(z_k|S_{k-1}^{k},z_{k-L}^{k-1})}
\ge 1 \bigg{\}}$. Now,
\begin{eqnarray*}
P(\hat{\bar{S}}|\bar{S}) &=& P(\hat{S}_0 \dots \hat{S}_{N-1}|S_0
\dots S_{N-1} )\\
&=& \int_{A(\bar{S}, \hat{\bar{S}})} \Pi_{k=0}^{N-1}f(z_k|S_{k-1}^{k},z_{k-L}^{k-1})d\bar{z} \\
& \le& \min_{\forall \rho_k }\int
\Pi_{k=0}^{N-1}f(z_k|S_{k-1}^{k},z_{k-L}^{k-1})
\\&\cdot& \Pi_{k=0}^{N-1}\bigg{(}\frac{f(z_k|\hat{S}_{k-1}^{k},z_{k-L}^{k-1})}{f(z_k|S_{k-1}^{k},z_{k-L}^{k-1})}\bigg{)}^{\rho_k}d\bar{z}\\
&=& \min_{\forall \rho_k}\int
\Pi_{k=0}^{N-1}(f(z_k|S_{k-1}^{k},z_{k-L}^{k-1}))^{1-\rho_k}
\\&\cdot& (f(z_k|\hat{S}_{k-1}^{k},z_{k-L}^{k-1}))^{\rho_k} d\bar{z}
\end{eqnarray*}


where $0\le \rho_k \le 1$ for $k = 0, \dots, N-1$.

The above integral can be simplified as follows.
\begin{align*}
&\int \Pi_{k=0}^{N-1}(f(z_k|S_{k-1}^{k},z_{k-L}^{k-1}))^{1-\rho_k}
\\& \cdot(f(z_k|\hat{S}_{k-1}^{k},z_{k-L}^{k-1}))^{\rho_k}
d\bar{z}\\
&= \int \Pi_{k=0}^{N-1} \frac{1}{\sqrt{2\pi}
\sigma^{1-\rho_k}(a_{k-I}^k) \hat{\sigma}^{\rho_k}(\hat{a}_{k-I}^k)}
\\& \cdot \exp (-\frac{(1-\rho_k)([-\bar{b}^T ~1
](z_{k-L}^k-\bar{\mu}(S_{k-1}^k)))^2 }{2\sigma^2(a_{k-I}^k)}
\\&
-\frac{\rho_k([-\bar{b}^T ~1
](z_{k-L}^k-\hat{\bar{\mu}}(\hat{S}_{k-1}^k)))^2
}{2\hat{\sigma}^2(\hat{a}_{k-I}^k)}) d\bar{z}\\
&= \Pi_{k=0}^{N-1} \int \frac{1}{\sqrt{2\pi}
\sigma^{1-\rho_k}(a_{k-I}^k) \hat{\sigma}^{\rho_k}(\hat{a}_{k-I}^k)}
\\& \cdot \exp (-\frac{(1-\rho_k)(u_k-\mathfrak{M}(S_{k-1}^k))^2
}{2\sigma^2(a_{k-I}^k)}\\&
-\frac{\rho_k(u_k-\hat{\mathfrak{M}}(\hat{S}_{k-1}^k))^2
}{2\hat{\sigma}^2(\hat{a}_{k-I}^k)}) du_k
\end{align*}
where $u_k=[-\bar{b}^T ~1 ]\cdot [z_{k-L}\dots z_k]^T$,
$\mathfrak{M}(S_{k-1}^k)) = [-\bar{b}^T ~1 ] \cdot
 \bar{\mu}(S_{k-1}^k)$,
$\hat{\mathfrak{M}}(\hat{S}_{k-1}^k))=[-\bar{b}^T ~1 ]\cdot
\hat{\bar{\mu}}(\hat{S}_{k-1}^k)$. The Jacobian matrix for the
change of variables has determinant equal to 1, since the
corresponding matrix of partial derivatives has ones on the diagonal
and is lower triangular. Note that the change of variables decouples
the original expression, so that it can be expressed as the product
of $N$ independent integrals. Now we can simplify the PEP as
follows.
\begin{align}
& P(\hat{\bar{a}}|\bar{a}) \le    \min_{\forall \rho_k }
\Pi_{k=0}^{N-1}\int \frac{1}{\sqrt{2\pi}
\sigma^{1-\rho_k}(a_{k-I}^k)
\hat{\sigma}^{\rho_k}(\hat{a}_{k-I}^k)} \nn \\
&\cdot \exp (-\frac{(1-\rho_k)(u_k-\mathfrak{M}(S_{k-1}^k))^2
}{2\sigma^2(a_{k-I}^k)} \nn
\\ &-\frac{\rho_k(u_k-\hat{\mathfrak{M}}(\hat{S}_{k-1}^k))^2
}{2\hat{\sigma}^2(\hat{a}_{k-I}^k)}) du_k \nn
\\
&=    \Pi_{k=0}^{N-1} \min_{\rho_k}\int \frac{1}{\sqrt{2\pi}
\sigma^{1-\rho_k}(a_{k-I}^k) \hat{\sigma}^{\rho_k}(\hat{a}_{k-I}^k)}
\nn
\\&\label{eq:errorboundintegral}\cdot \exp
(-\frac{(1-\rho_k)(u_k-\mathfrak{M}(S_{k-1}^k))^2
}{2\sigma^2(a_{k-I}^k)}
\nn\\&-\frac{\rho_k(u_k-\hat{\mathfrak{M}}(\hat{S}_{k-1}^k))^2
}{2\hat{\sigma}^2(\hat{a}_{k-I}^k)}) du_k\\
&= \Pi_{k=0}^{N-1} \min_{ \rho_k} \frac{\sigma^{\rho_k}(a_{k-I}^k)
\hat{\sigma}^{1-\rho_k}(\hat{a}_{k-I}^k)}{\sqrt{(1-\rho_k)\hat{\sigma}^2(\hat{a}_{k-I}^k)
+ \rho_k\sigma^2(a_{k-I}^k) }}\nn \\&\cdot \exp (
-\frac{(1-\rho_k)\hat{\sigma}^2(\hat{a}_{k-I}^k)\mathfrak{M}^2(S_{k-1}^k)
+
\rho_k\sigma^2(a_{k-I}^k)\hat{\mathfrak{M}}^2(\hat{S}_{k-1}^k)}{2\sigma^2(a_{k-I}^k)\hat{\sigma}^2(\hat{a}_{k-I}^k)}
\nn \\ & +
\frac{((1-\rho_k)\hat{\sigma}^2(\hat{a}_{k-I}^k)\mathfrak{M}(S_{k-1}^k)
+
\rho_k\sigma^2(a_{k-I}^k)\hat{\mathfrak{M}}(\hat{S}_{k-1}^k))^2}{2\sigma^2(a_{k-I}^k)\hat{\sigma}^2(\hat{a}_{k-I}^k)((1-\rho_k)\hat{\sigma}^2(\hat{a}_{k-I}^k)
+ \rho_k\sigma^2(a_{k-I}^k))}) \nn \\\label{eq:errorbound}& =
\Pi_{k=0}^{N-1} W(S_{k-1}^k,\hat{S}_{k-1}^k)
\end{align}
where $W(S_{k-1}^k,\hat{S}_{k-1}^k)$ is a function of
$\sigma(a_{k-I}^k)$, $\hat{\sigma}(\hat{a}_{k-I}^k)$,
$\mathfrak{M}(S_{k-1}^k)$ and $\hat{\mathfrak{M}}(\hat{S}_{k-1}^k)$
and the simplification of the integral in
(\ref{eq:errorboundintegral}) is given in the Appendix.

It is important to note that the factorization of PEP given by
(\ref{eq:errorbound}) for our channel model is possible because the
autoregressive filter $\bar{b}$ is not dependent on the input bit
sequence. In \cite{kavcic2000}, $\bar{b}$ is assumed to be data
dependent given by $\bar{b}(a_{k-I}^k)$. When the autoregressive
filter $\bar{b}(a_{k-I}^k)$ becomes data dependent, it is very
difficult to write PEP in the form given in (\ref{eq:errorbound}).
In this case, the inverse of the autoregressive filter of the noise
process $n_k$ ($[-\bar{b}(a_{k-I}^k)^T ~1 ]$) is state-dependent
which means that the actual state transition ($S_{k-1}^{k}$) and
estimated state
transition ($\hat{S}_{k-1}^k$) have different filters. In this situation, the specific change of variables used above does not seem to work.

 Probability of bit error can now be
further simplified as \cite{biglieri84},
\begin{align*}
&P_b(e) \le \sum_{N=1}^{\infty} \sum_{\bar{S}} P (\bar{S})
\sum_{\hat{\bar{S}}: ~(\bar{S},\hat{\bar{S}}) \in E_N } \frac{d}{dZ}
\\& ~~~~~~~~~~ \cdot \big{[} \Pi_{k=0}^{N-1} Z^{\delta(a_k,\hat{a}_k)}|_{Z=1}
P(\hat{\bar{S}}|\bar{S}) \big{]}
\\& =\frac{d}{dZ}
\sum_{N=1}^{\infty} \frac{1}{M}.\frac{1}{2^{N}} \sum_{\bar{S}}
\sum_{\hat{\bar{S}}} \Pi_{k=0}^{N-1} Z^{\delta(a_k,\hat{a}_k)}
P(\hat{\bar{S}}|\bar{S}))|_{Z=1}
\end{align*}
\begin{align*}
& \le \frac{d}{dZ} \sum_{N=1}^{\infty} \frac{1}{M}.\frac{1}{2^{N}}
\sum_{\bar{S}} \sum_{\hat{\bar{S}}} \Pi_{k=0}^{N-1}
Z^{\delta(a_k,\hat{a}_k)} \\
& ~~~~\cdot W(S_{k-1}^k,\hat{S}_{k-1}^k))|_{Z=1} ~~\mbox{(Using
(\ref{eq:errorbound}))}
\\&
= \frac{1}{M}\frac{d}{dZ} \sum_{N=1}^{\infty}  \sum_{\bar{S}}
\sum_{\hat{\bar{S}}} \Pi_{k=0}^{N-1}
\frac{1}{2}Z^{\delta(a_k,\hat{a}_k)}
W(S_{k-1}^k,\hat{S}_{k-1}^k))|_{Z=1}
\\&
=\frac{1}{M}\frac{d}{dZ}(T(Z))|_{Z=1}
\end{align*}
For obtaining $T(Z)$, we construct a product trellis. Consider a
matrix $V(Z)$ of order $M^2\times M^2$, where each row and column is
indexed by a pair of states corresponding to the actual and the
errored states. Let $\beta_i$ represent a state that takes one of
$2^{L+I}$ values. Consider the entry of $V(Z)$ indexed by
$((\beta_i, \beta_j), (\beta_i', \beta_j'))$,

\begin{align*}
&[V(Z)]_{((\beta_i, \beta_j), (\beta_i', \beta_j')}\\& =
\begin{cases}
 \frac{1}{2}Z^{\delta(a_i,a_j)}
W((\beta_i,
\beta_j), (\beta_i', \beta_j'))\\
0, ~\mbox{if either $\beta_i\rightarrow\beta_i'$ or
$\beta_j\rightarrow \beta_j'$ not allowed}
\end{cases}
\end{align*}
where $W((\beta_i, \beta_j), (\beta_i', \beta_j'))$ can be found for
state transitions $\beta_i\rightarrow\beta_i'$ and
$\beta_j\rightarrow \beta_j'$ using (\ref{eq:errorbound}) and $a_i$
and $a_j$ are latest bit in states $\beta_i'$ and $\beta_j'$
respectively. A product state is called good state if
$\beta_i'=\beta_j'$ and bad otherwise. $V(Z)$ will have a structure
which has $V_{GG}(Z)$ (good to good state transition), $V_{GB}(Z)$
(good to bad state transition), $V_{BG}(Z)$ (bad to good state
transition) and $V_{BB}(Z)$ (bad to bad state transition),
\begin{eqnarray*}
V(Z)=\left[\begin{array}{cc} V_{GG}(Z) & V_{GB}(Z)\\
V_{BG}(Z) & V_{BB}(Z)\end{array}\right]
\end{eqnarray*}
where the order of $V_{GG}(Z)$ matrix is $M\times M$ and the order
of $V_{BB}(Z)$ is $(M^2-M) \times (M^2-M)$. Now we can write $T(Z)$
as,
\begin{eqnarray*}
T(Z) = \mathfrak{a}(Z) +
\mathfrak{b}(Z)(I-V_{BB}(Z))^{-1}\mathfrak{c}(Z)
\end{eqnarray*}
where $\mathfrak{a}(Z)= \mathds{1}^T V_{GG}(Z) \mathds{1}$,
$\mathfrak{b}(Z)=\mathds{1}^T V_{GB}(Z)$ and
$\mathfrak{c}(Z)=V_{BG}(Z) \mathds{1}$. The symbol $\mathds{1}$
denotes a vector all of whose entries are 1 and $I$ is identity
matrix of order $(M^2-M) \times (M^2-M)$. Using the above result, we
can compute $P_b(e)$ as \cite{biglieri84},
\begin{align*}
& P_b(e) \le \frac{1}{M}
[\mathfrak{a}'(1)+\mathfrak{b}'^T(1)(I-V_{BB}(1))^{-1}\mathfrak{c}(1)+\mathfrak{b}^T(1)\\
&\cdot (I-V_{BB}(1))^{-1}\mathfrak{c}'(1)
+\mathfrak{b}^T(1)(I-V_{BB}(1))^{-1}\\& \cdot
V'_{BB}(1)(I-V_{BB}(1))^{-1}\mathfrak{c}(1)].
\end{align*}

\noindent For our model, $V_{GG}(Z)$ is not a function of $Z$ which
means that $\mathfrak{a}'(1)=0$. Similarly, $\mathfrak{c}(Z)$ is
also not a function of $Z$ which implies $\mathfrak{c}'(1)=0$ and it
should also be noted that $\mathfrak{b}'^{T}(1)= \mathfrak{b}^T(1)$.
The new bound for our channel model is,
\begin{align*}
&P_b(e) \le \frac{1}{M}
[\mathfrak{b}^T(1)(I-V_{BB}(1))^{-1}\mathfrak{c}(1)+\mathfrak{b}^T(1)\\
&\cdot
(I-V_{BB}(1))^{-1}V'_{BB}(1)(I-V_{BB}(1))^{-1}\mathfrak{c}(1)].
\end{align*}

\vspace{-.3cm}
\section{Simulation Results}\label{sec:Simulation Results}

\begin{figure}
\centering
\includegraphics[width=3.5in]{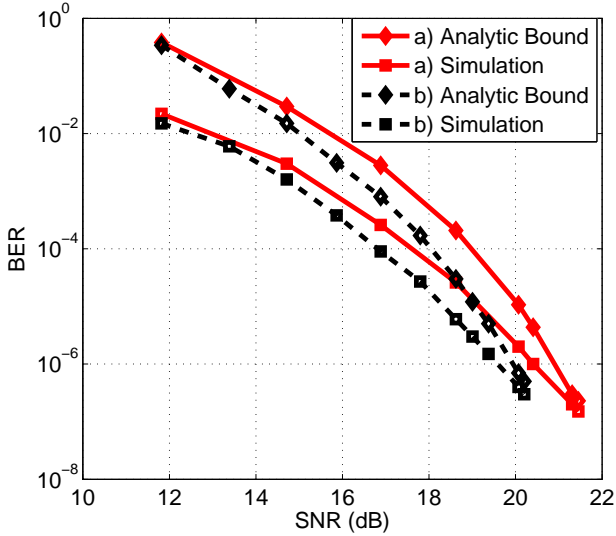}
\caption{\label{fig:results1}  BER with different SNR for the
channel model with a) $8$ states in decoding and b) 16 states in
decoding.}
\end{figure}
In the first set of simulations, we used the following parameters:
$L=2$ with $\bar{b}=[.1 ~.5]$ and ISI memory $I = 1$. The signal
dependent noise variance for $4$ states are given by
$\sigma^2(00)=1$, $\sigma^2(01)=2$, $\sigma^2(10)=3$ and
$\sigma^2(11)=4$. The number of states in decoding is equal to $8$
in this case. The SNR is defined as signal energy in $y(a_{k-I}^k)$
divided by total noise variance. We have used a linear signal
component given as $y(a_{k-1}^{k})=c(2a_k+a_{k-1})$ where the value
of $c$ can be varied to change the SNR. In Figure
\ref{fig:results1}, the analytic bound follows the simulation BER.
At an SNR of $21$ dB, the analytic bound gives a BER equal to
$3\times 10^{-7}$ whereas simulation BER is equal to $2\times
10^{-7}$. The analytic bound is quite tight in high SNR regime. In
another simulation, we used following parameters, $L=3$ coefficients
of an autoregressive filter is given by $\bar{b}=[.1 ~.3~.5]$, ISI
memory ($I$) is equal to $1$ and signal dependent noise variance for
$4$ states are given by $\sigma^2(00)=1$, $\sigma^2(01)=2$,
$\sigma^2(10)=3$ and $\sigma^2(11)=4$. The number of states in
decoding is equal to $16$ in this case. In Figure
\ref{fig:results1}, the analytic bound again follows the simulation
BER for this channel model with modified channel parameters. At an
SNR of $20$ dB, the analytic bound gives a BER equal to $7\times
10^{-7}$ whereas simulation BER is equal to $4\times 10^{-7}$.

\section{Conclusions and future work}\label{sec:conclusions}
We considered the problem of deriving an analytical upper bound for
ML sequence detection in ISI channels with signal dependent
Gauss-Markov noise. In these channels the pairwise error probability
(PEP) is not symmetric. Moreover, it is hard to express the PEP as a
product of appropriate terms that allow the application of flowgraph
techniques. In this work, we considered a subset of these channels,
and demonstrated an appropriate upper bound on the PEP. Using this
upper bound along with pairwise state diagrams, we arrive at
analytical BER bounds that are tight in the high SNR regime. These
bounds have been verified by our simulation results.

It would be interesting to examine whether our current techniques
can be extended to address the general channel model. Moreover, it
may be possible to reduce the complexity of evaluating the bound by
reducing the size of the product trellis by exploiting channel
characteristics. We are currently investigating these issues.

\begin{appendix}
The integral in the equation (\ref{eq:errorboundintegral}) can be
expressed in the following form,
\begin{align*}
&\int \frac{1}{\sqrt{2\pi}\gamma} \exp (-\frac{1}{2}
(\alpha(x-m)^2+\beta(x-\hat{m})^2))dx\\
&=\frac{1}{\gamma\sqrt{\alpha+\beta}} \exp(-\frac{\alpha m^2+\beta
\hat{m}^2}{2}+ \frac{(\alpha m +\beta
\hat{m})^2}{2(\alpha+\beta)})\\
&\cdot \underbrace{\int\frac{\sqrt{\alpha+\beta}}{\sqrt{2\pi}}
\exp(-\frac{\alpha+\beta}{2}(x-\frac{\alpha m +\beta
\hat{m}}{\alpha+\beta})^2)dx}_{= 1}
\\&=\frac{1}{\gamma\sqrt{\alpha+\beta}} \exp(-\frac{\alpha m^2+\beta
\hat{m}^2}{2}+ \frac{(\alpha m +\beta \hat{m})^2}{2(\alpha+\beta)})
\end{align*}
where $\alpha=\frac{(1-\rho_k)}{\sigma^2(a_{k-I}^k)}$,
$\beta=\frac{\rho_k}{\hat{\sigma}^2(\hat{a}_{k-I}^k)}$,
$m=\mathfrak{M}(S_{k-1}^k)$, $\gamma=\sigma^{1-\rho_k}(a_{k-I}^k)
\hat{\sigma}^{\rho_k}(\hat{a}_{k-I}^k)$ and
$\hat{m}=\hat{\mathfrak{M}}(\hat{S}_{k-1}^k)$. Using the above
equality, we can easily simplify the RHS of equation
(\ref{eq:errorboundintegral}) as,
\begin{align*}
&\frac{1}{\sigma^{1-\rho_k}(a_{k-I}^k)
\hat{\sigma}^{\rho_k}(\hat{a}_{k-I}^k)} \cdot\frac{\sigma(a_{k-I}^k)
\hat{\sigma}(\hat{a}_{k-I}^k)}{\sqrt{(1-\rho_k)\hat{\sigma}^2(\hat{a}_{k-I}^k)
+ \rho_k\sigma^2(a_{k-I}^k) }}\nn \\&\cdot \exp (
-\frac{(1-\rho_k)\hat{\sigma}^2(\hat{a}_{k-I}^k)\mathfrak{M}^2(S_{k-1}^k)
+
\rho_k\sigma^2(a_{k-I}^k)\hat{\mathfrak{M}}^2(\hat{S}_{k-1}^k)}{2\sigma^2(a_{k-I}^k)\hat{\sigma}^2(\hat{a}_{k-I}^k)}
\nn \\ & +
\frac{((1-\rho_k)\hat{\sigma}^2(\hat{a}_{k-I}^k)\mathfrak{M}(S_{k-1}^k)
+
\rho_k\sigma^2(a_{k-I}^k)\hat{\mathfrak{M}}(\hat{S}_{k-1}^k))^2}{2\sigma^2(a_{k-I}^k)\hat{\sigma}^2(\hat{a}_{k-I}^k)((1-\rho_k)\hat{\sigma}^2(\hat{a}_{k-I}^k)
+ \rho_k\sigma^2(a_{k-I}^k))}). \nn \\
\end{align*}
\end{appendix}
\vspace{-1cm}

\bibliographystyle{plain}

\end{document}